# Plasmonic sensor based on the Ebbesen effect

P.N. Melentiev,[1,2,*] A.S. Gritchenko,[1,2] A.S. Baburin,[3,4] N.A. Orlikovsky,[3,4] A.A. Dobronosova,[3,4] I.A. Rodionov,[3,4] V.I. Balykin[1,2]

[1] *Institute of Spectroscopy RAS, Troitsk, Moscow 108840, Russia*
[2] *National Research University, Higher School of Economics, Moscow 101000, Russia*
[3] *Bauman Moscow State Technical University, Moscow 105005, Russia*
[3] *Dukhov Research Institute of Automatics, Moscow 127055, Russia*
*Corresponding author: melentiev@isan.troitsk.ru*

**We present a new method for measuring low concentrations and simultaneously small volumes of fluorescent molecules based on the use of the Ebbesen effect of the extraordinary transmission (EOT) of light through an array of nano-holes. In the method the EOT effect is realized at the fluorescence wavelength of the detected molecules with a low transmission of light at the absorption wavelength. The approach allows realizing high level of the sensor sensitivity due to suppression of the inevitable parasitic luminescence of the sensor substrate. The method was demonstrated by detecting an ultra low concentration (at a level of 20 pg/ml (3 p.p.t.)) and an ultra-small volume (about 5 μl) of Cy-5 fluorescent markers in a dimethyl sulfoxide solution.**

Small concentration detection is extremely important both in fundamental research and in various applications. The ability to detect a single atom or molecule is a key aspect of studying the single quantum object properties and QED effects [1-5]. The practical importance of ultra low concentration detection at a level of single molecules is the clinically relevant biomarkers detection in blood serum [6-8].

The most developed methods of single molecule detection are based on their fluorescent measurements [3,4,6-9]. However, the practical implementation of single-molecule sensitivity, as a rule, is limited by the background parasitic signal [3,4]. Such background signal has different origins: (1) scattering of laser light, (2) luminescence of various materials (substrates, solvents). As a result it is hard to eliminate the luminescence of media surrounding the detected molecule from a molecule fluorescence leading to the known trade-off between a volume of molecules excitation (a large volume is needed to perform fast measurements) and sensitivity of corresponding single molecules counting (SMC) sensors (a small volume reduces parasitic luminescence helping realization a single molecule detection) [6].

In this work we present a method based on the use of the EOT light transmission through an array of nano-holes perforated in a metal film (plasmonic crystal, PC) that allows suppressing the various types of parasitic luminescence and having sensitivity approaching sensitivity of the SMC sensors. We develop a new type of sensor based on (1) optical excitation of molecules, (2) their subsequent fluorescence detection and (3) the use of the Ebbesen effect of the extraordinary transmission of light (EOT).

The EOT effect is already used in sensorics [10]. It was shown that small variations in the refractive index (RI) of media adjacent to nanoholes array (plasmonic crystal) lead to changes in the dispersion of excited plasmonic waves, which in turn results in shifting of the resonant wavelength of the EOT effect measured through an optical transmission of the nanoholes array [11-16]. Note that the sensors based on such a use of the EOT are characterized by a moderate sensitivity and in principle can't reach the SMC sensors sensitivity, since the RI change due to a single molecule presence is negligible compared to the RI changes caused by the temperature or the pressure instabilities. The minimal detecting RI change corresponds to $10^8$ molecules introduced in the sensor detection region [17].

In this work, we utilize the EOT effect in a completely new way to perform effective fluorescence measurements: to create an optical filter, which allows a significant increase in the signal-to-noise ratio, due to a significant decrease in parasitic substrate luminescence. The spectral transmission of the PC (nanoholes array) is chosen in such a way: (1) to achieve high transmission of light (the EOT effect) at the wavelength of the molecule fluorescence and, (2) to achieve low light transmission at the laser wavelength. In such approach one can significantly suppress parasitic luminescence of the substrate caused by the exciting laser light, which is usually the main limitation factor of sensor sensitivity based on the fluorescence measurements of arbitrary large volumes of analyte, namely more than 1 nl. Note that the sensitive fluorescence measurements of such arbitrary large volumes are a key factor in reaching fast measurements of SMC sensors [6].

We note as well that such sensor allows further increasing the sensitivity through another effect. Fabrication of high quality PC helps to substantially reduce large background signal accompanying sensing through a transparent dielectric substrate. The large background signal is formed by a laser scattering on the quartz surface and the particles luminescence on the surface. These surface imperfections have different origins: surface polishing defects, adsorption of particles from air, etc. The detection of molecules fluorescence through a high quality PC has two main advantages. First, luminescence from such particles does not enter the detection zone of fluorescence, since the microparticles block the passage of light through the nanohole. Second, most of the particles are located between nano-holes, and their luminescence does not reach the detection zone.

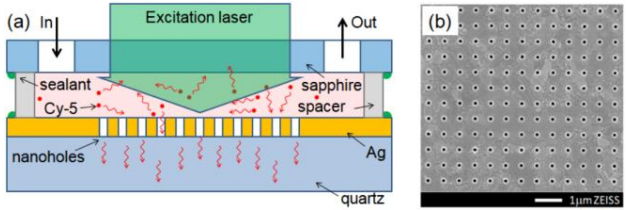

Fig. 1. The plasmonic sensor based on the Ebbesen effect: (a) a schematic diagram of the sensor, (b) an electron microscope image of a PC formed by a matrix of nano-holes perforated in a 100 nm thick silver film.

The EOT transmission through a PC formed by nanoholes perforated in metal film consists in the energy transfer of the incident light wave through the PC to its other side via surface plasmons [11,18,19]. The light transmission through the PC can be much higher than, so-called, geometric transmission, determined by the total area of all the holes of the PC.

At normal illumination of a PC, the wave vector of the excited plasmon wave is determined by the following expression [18]:

$$\vec{k}_{SPP} = \left(\frac{2n_x\pi}{\Lambda}\right)\vec{i} + \left(\frac{2n_y\pi}{\Lambda}\right)\vec{j}$$

Here $\Lambda$ is the pitch of nano-holes array, $n_x$ and $n_y$ are integers. Thus, spectral regions of the EOT effect and low light transmittance of a PC are realized at the wavelengths determined by the period of the nano-hole array [18,19].

**Fig. 1a** illustrates the design of a plasmonic sensor. The senosor is built in such a way to effectively measure fluorescence of a single drop of Cy-5 molecules in a solution. The sensor is based on the use of ultra-high quality PC [20-22], formed by the nanoholes of 175 nm diameter perforated in 100 nm thick large grains silver film [23,24] on a quartz substrate (Fig. 1b). The size of the nano-holes array is 1 × 1 mm. In the sensor, we used an analyte solution (Cy-5 dye molecule in dimethyl sulfoxide, DMSO) placed between two surfaces: (1) a PC surface and (2) a surface of a YAG substrate with a low surface roughness (Shinkosha Ltd.). In the YAG substrate the two 2 mm diameter holes were formed for input and output of the analyte. Between these surfaces the quartz stripes were arranged to create a 100 μm thick gap. The listed components of the plasmonic sensor are glued together, which gives the sensor a mechanical stability, tightness and ease of use.

The volume of the plasmonic sensor which is filled by analyte is equal to 5 μL. A drop of analyte solution is introduced into the sensor using a pipetator through one of the holes made in the YAG substrate. Further, by the forces of surface tension, this drop itself is dragged into the sensor, and it is spread evenly without bubbles over the entire surface of the PC.

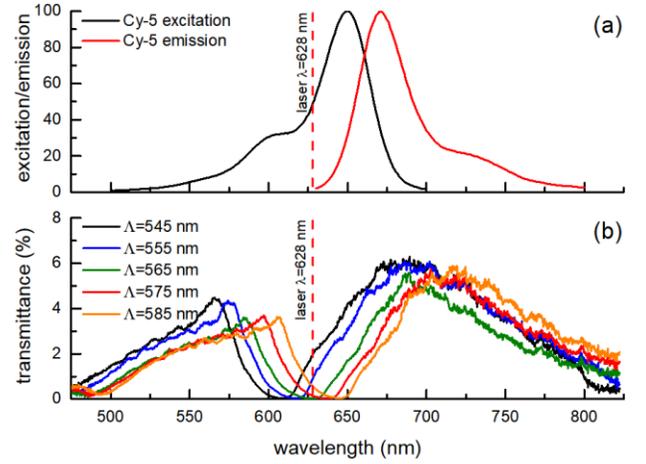

Fig. 2. Spectral selectivity of a plasmonic sensor based on the Ebbesen effect: (a) excitation spectra of Cy-5, (b) transmission spectra of nanoholes arrays having different pitch $\Lambda$. The dashed line indicates the laser wavelength $\lambda$ = 628 nm.

We note that a detailed theoretical analysis of the plasmonic sensor under investigation is a rather difficult problem, requiring consideration of a number of effects: (1) the excited spatial hybrid "photon + plasmon" modes [20], (2) changes in the refractive index of DMSO when Cy-5 molecules are introduced into it [25], (3) the adsorption of Cy-5 molecules on the surface [26], (4) the change in the transmission of light through the nanoholes when the Cy-5 molecules are close to the nanohole [27], (5) the QED effects in the fluorescence of molecules near the nanohole [28].

The plasmonic sensor was located in the object plane of the Nikon Eclipse/Ti-U microscope. The sensor is illuminated by laser radiation at wavelength $\lambda$ = 628 nm and a power of 10 mW from a diode laser. The laser radiation illuminates the surface of the PC normally to the Ag film. The size of the laser spot on the sample equals to 20 μm.

The fluorescence of Cy-5 molecules excited by laser radiation is passed through the nanohole array and collected with a ×10 microscope objective on the EM CCD camera (Princeton Instruments). The CCD camera is installed next to a bandpass filter with a bandwidth of 100 nm, corresponding to the center of the fluorescence wavelength of the Cy-5 dye (see **Fig. 2a**).

**Fig. 2b** presents the measured transmittance spectra of various sensors formed by PCs with different pitches of nanoholes arrays $\Lambda$: 545 nm, 555 nm, 565 nm, 575 nm and 585 nm. This figure clearly shows spectral windows of high and low transmission of light due to the EOT effect. During these measurements, the sensors were filled with a pure DMSO solution without Cy-5 molecules. As it can be seen from the figure, the transmission spectra are characterized by a region of high and low transmittance. The spectral position of these regions depends on the nanoholes arrays pitch $\Lambda$.

The smallest transmission at the wavelength $\lambda$ = 628 nm is realized through a PC with a nanoholes array pitch $\Lambda$ = 565 nm.

The measured transmittance equals to $T_{exc}$ = 0.03% (PC transmittance at the wavelength of laser light). Such a strong attenuation of laser radiation by a PC makes it possible to significantly weaken the luminescence of the quartz substrate excited by laser radiation.

The highest transmission of a PC with $\Lambda$ = 565 nm is realized at wavelength of the Cy-5 molecules and is equal to $T_{Cy-5}$ = 5% (PC transmittance at the wavelength of Cy-5 molecules luminescence). Thereby, the sensitivity increase realized by the PC is characterized by the figure of merit FOM = $T_{Cy-5} / T_{exc} \approx 166$. Note that such a high FOM value is a consequence of the high quality of the PC formed by: (i) silver film having minimal losses for plasmon waves, (ii) identical size and shape of nanoholes, (iii) identity of the distance between nanoholes on the entire area of the PC.

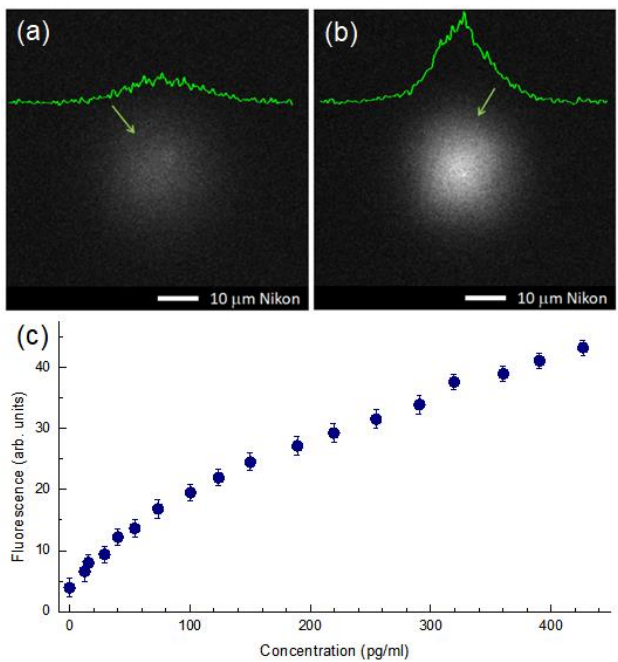

Fig. 3. Detection of Cy-5 molecules in the DMSO solution with the plasmonic sensor: (a) measured by 2D CCD camera luminescence of a pure DMSO solution without Cy-5 molecules in the plasmonic sensor; (b) the measured luminescence of a Cy-5 molecules solution having concentration n = 40 pg/ml in the plasmonic sensor; (c) measured by 2D CCD camera amplitudes of luminescence of different concentrations of a Cy-5 molecules solution introduced in the plasmonic sensor. The light-green curves in Figures (a) and (b) show sections of optical images.

**Fig. 3** presents the results of measurements of the fluorescence of Cy-5 molecules dissolved in DMSO at different concentrations and clearly shows that fluorescence measurements with the use of the EOT effect can realize detection of ultra low concentrations of Cy-5 molecules.. **Fig. 3a** shows an optical image of the PC surface on a CCD camera in the case of a pure DMSO solution (without Cy-5 molecules) in the sensor. As can be seen in the figure, the optical image shows a white spot corresponding to the luminescence of the quartz substrate used to build the sensor. The green curve in the graph corresponds to the image cut of the spot along the horizontal axis. The amplitude of this parasitic luminescence determines the minimum detectable concentration of dye molecules by the sensor.

When a solution of Cy-5 dye molecules with a concentration of 40 pg/ml is introduced into the sensor, a noticeable increase in the recorded signal occurs in the image on the CCD camera (**Fig. 3b**). Comparison of the images in **Fig. 3a,b** shows a three-fold increase of the signal when a solution of the dye molecules is introduced.

**Fig. 3c** presents detection of various concentrations of Cy-5 molecules in a DMSO solution. From this figure it can be seen that the measured fluorescence signal directly depends on the concentration of the Cy-5 molecules in the solution and the plasmonic sensor allows one to confidently register the dye molecules in the range from 20 pg/ml to 400 pg/ml.

The minimum recorded concentration of dye molecules is 20 pg/ml (3 p.p.t.). This is the concentration at which the detected fluorescence signal from the molecules is twice as high as the level of parasitic luminescence. At such a low concentration of Cy-5 the distance between molecules is about 3.5 µm. With the chosen laser beam sizes, this concentration corresponds to detection of only about $10^3$ Cy-5 molecules in the entire volume of the plasmonic sensor (compared to about $10^8$ molecules for RI sensors), which corresponds to detection of the total dye molecules mass less than 1 atto gram which is a new record in plasmonic sensor sensitivity [17,29]. Note that the sensitivity is approaching the detection limit of SMC sensors, since for the SMC sensors it is needed to detect more than $10^3$ molecules to exceed a shot noise due to the Poisson molecule sampling [6]. The SMC sensors demands hours to detect statistically reliable data [6], while our approach helps to detect the same number of molecules during a single short of laser light.

In the investigated range of concentrations of dye molecules, the detected fluorescence signal should be proportional to the number of molecules. However, as can be seen from **Fig. 3c**, the measured fluorescence dependence of Cy-5 molecules has a much more complex dependence. We explain the distinction by the quenching of the molecules fluorescence when they form agglomerates, since the number of agglomerated molecules directly depends on their concentration [30].

The Cy-5 dye molecules used in the plasmonic sensor are known as fluorescent biomarkers [31]. These molecules can be attached through appropriate antibodies to various biomolecules which are serving as markers of various processes in the human body [32]. Thus, the demonstrated plasmonic sensor has a potential to become a hardware platform for the medical applications using small volume of blood plasma which has to be analyzed at the highest sensitivity level competitive with SMC sensors.

Summarizing, we have developed plasmonic sensor of fluorescent biomarkers Cy-5, based on the use of Cy-5 fluorescence measurements through a PC having high level of spectral selectivity realized by the EOT effect. A record level of sensitivity is shown with the use of plasmonic structures pawing a way to new types of ultrasensitive plasmonic sensors.

**Funding.** Russian Foundation for Basic Research (17-02-01093); Advanced Research Foundation (contract number 7/004/2013-2018 on 23.12.2013).

**Acknowledgment**. We thank V.V. Klimov for helpful discussions. Samples were made at the BMSTU Nanofabrication Facility (Functional Micro/Nanosystems, FMNS REC, ID 74300).

## References

1. K. Kneipp, Y. Wang, H. Kneipp , L. T. Perelman, I. Itzkan, R. R. Dasari, and M. S. Feld, Phys. Rev. Lett. **78**, 1667 (1997).
2. C. Zander , J. Enderlein, and R. A. Keller, in Single Molecule Detection in Solution: Methods and Applications, C. Zander, J. Enderlein, R. A. Keller, eds. (Wiley-VCH, 2002), pp. 386.
3. K. P. Nayak, P. N. Melentiev, M. Morinaga, F. L. Kien, V. I. Balykin, and K. Hakuta, Opt. Express **15**, 5431 (2007).
4. A. V. Akimov, A. Mukherjee, C. L. Yu, D. E. Chang, A. S. Zibrov, P. R. Hemmer, H. Park and M. D. Lukin, Nature **450**, 402 (2007)
5. V.I. Balykin, P.N. Melentiev, Phys. Usp., **61**, 133 (2018)
6. D.R. Walt  Anal. Chem. **85**, 1258-1263 (2013).
7. F. Ma, Y. Li, B. Tang, and C.Y. Zhang, Acc. Chem. Res. 49, 1722-1730 (2016).
8. A.B. Taylor, and P. Zijlstra, ACS Sensors **2**, 1103-1122 (2017).
9. R. Mitsch, C. Sayrin, B. Albrecht, P. Schneeweiss, and A. Rauschenbeutel, Nat. Commun. **5**, 5713 (2014).
10. T. W. Ebbesen, H. J. Lezec, H. F. Ghaemi, T. Thio, and P. A . Wolff, Nature **391**, 667 (1998).
11. K. L. Van der Molen, F. B. Segerink, N. F. Van Hulst, and L. Kuipers, Appl. Phys. Lett. **85**, 4316 (2004).
12. A. G. Brolo, R. Gordon, B. Leathem, and K. L. Kavanagh, Langmuir **20**, 4813 (2004).
13. R. Gordon, D. Sinton, K. L. Kavanagh, and A. G. Brolo, Acc. Chem. Res. **8**, 1049 (2008).
14. Genet, C., and T. W. Ebbesen, Nature **445**, 39 (2007).
15. A. G. Brolo, E. Arctander, R. Gordon, B. Leathem, and K. L. Kavanagh, Nano Lett. **4**, 2015 (2004).
16. S. Roh, T. Chung, and B. Lee, Sensors **11**, 1565 (2011).
17. V.N. Konopsky, and E.V. Alieva, Anal. Chem. **79**, 4729 (2007).
18. H. F. Ghaemi, T. Thio, D. E. Grupp, T. W. Ebbesen, and H. J. Lezec, Phys. Rev. B **58**, 6779 (1998).
19. L. Martin-Moreno, F. J. Garcia-Vidal, H. J. Lezec, K. M. Pellerin, T. Thio, J. B. Pendry, and T. W. Ebbesen, Phys. Rev. Lett. **86**, 1114 (2001).
20. P. Melentiev, A. Kalmykov, A. Gritchenko, A. Afanasiev, V. Balykin, A. Baburin, E. Ryzhova, I. Filippov, I. Rodionov, I. Nechepurenko, A. Dorofeenko, I. Ryzhikov, A. Vinogradov, A. Zyablovsky, E. Andrianov, and A.A. Lisyansky, Appl. Phys. Lett. **111**, 213104 (2017).
21. I. A. Rodionov, A. S. Baburin, A. V. Zverev, I. A. Philippov, A. R. Gabidulin, A. A. Dobronosova, E. V. Ryzhova, A. P. Vinogradov, A. I. Ivanov, S. S. Maklakov, A. V. Baryshev, I. V. Trofimov, A. M. Merzlikin, N. A. Orlikovsky, and I. A. Rizhikov, Proc. SPIE **10343**, 1034337 (2017).
22. A. S. Baburin, A. I. Ivanov, I. V. Trofimov, A. A. Dobronosovaa, P. N. Melentiev, V. I. Balykin, D. O. Moskalev, A. A. Pishchimova, L. A. Ganieva, I. A. Ryzhikov, and I. A. Rodionov, Proc. SPIE **10672**,  106724D (2018).
23. A. S. Baburin, A. R. Gabidullin, A. V. Zverev, I. A. Rodionov, I. A. Ryzhikov, and Y. V. Panfilov, Vestn. Mosk. Gos. Tekh. Univ. im. N.E. Baumana, Priborostr. [Herald of the Bauman Moscow State Tech. Univ., Instrum. Eng.], **6**, 4–14 (2016)
24. A. S. Baburin, A. I. Ivanov, I. A. Ryzhikov, I. V. Trofimov, A. R. Gabidullin, D.O. Moskalev, Y. V. Panfilov, and I. A. Rodionov, in Proceedings of IEEE Progress In Electromagnetics Research Symposium — Spring (PIERS) (IEEE, 2017), pp. 1497-1502.
25. C. Genet, and T. W. Ebbesen, Nature **445**, 39 (2007).
26. E.G. McRae, and M. Kasha, J. Chem. Phys. **28**, 721 (1958).
27. R.M. Gelfand, S. Wheaton, and R. Gordon, Opt. Lett. **39**, 6415 (2014).
28. A.E. Afanasiev, P.N. Melentiev, A.A. Kuzin, A.Yu. Kalatskiy and V.I. Balykin, New J. Phys. **18**, 053015 (2016).
29. X. Fan, I. M. White, S. I. Shopova, H. Zhu, J. D. Suter, and Y. Sun, Anal. Chim. Acta **620**, 8 (2008).
30. P. J. Hillson, and R. B. McKay, T. Faraday. Soc. **61**, 374 (1965).
31. Y. Liu, J. Bishop, L. Williams, S. Blair, and J. Herron, Nanotechnology **15**, 1368 (2004).
32. Y. Sandoval, S.W. Smith, A.S.V. Shah, A. Anand, A.R. Chapman, S.A. Love, K. Schulz, J. Cao, N.L. Mills, and F.S. Apple, Clin. Chem. **63**, 369 (2017).